\documentclass[useAMS,usenatbib]{mn2e}
\usepackage{graphicx}
\usepackage{txfonts}
\usepackage{natbib}

\title[Optical polarization in EXO\,2030+375]{The high optical polarization
in the Be/X-ray binary EXO 2030+375}
\author[P. Reig et al.]{P.~Reig$^{1,2}$\thanks{E-mail: pau@physics.uoc.gr},
    D. Blinov$^{2,3}$, I. Papadakis$^{2,1}$, N. Kylafis$^{2,1}$, K. Tassis$^{2,1}$	\\
$^{1}$ IESL, Foundation for Research and Technology-Hellas, GR-71110 Heraklion,
Crete, Greece\\
$^{2}$ Department of Physics \& Institute of Theoretical \& Computational 
Physics, University of Crete, PO Box 2208, GR-710 03, Heraklion, Crete,
Greece \\
$^{3}$ Astronomical Institute, St. Petersburg State University,Universitetsky pr. 28,
 Petrodvoretz, 198504 St. Petersburg, Russia}

\newcommand{\exo}  {EXO\,2030+375}

\def\simless{\mathbin{\lower 3pt\hbox
     {$\rlap{\raise 5pt\hbox{$\char'074$}}\mathchar"7218$}}}   
\def\simmore{\mathbin{\lower 3pt\hbox
     {$\rlap{\raise 5pt\hbox{$\char'076$}}\mathchar"7218$}}}   

\def\msun{~{\rm M}_\odot}
\def\rsun{~{\rm R}_\odot}

\begin{document}

\date{Accepted ??. Received ??; in original form ??}

\pagerange{\pageref{firstpage}--\pageref{lastpage}} \pubyear{2012}

\maketitle

\label{firstpage}

\begin{abstract}

Polarization in classical Be stars results from Thomson scattering of the
unpolarized light from the Be star in the circumstellar disc. Theory and
observations agree that the maximum degree of polarization  from isolated
Be stars is $\simless4$\%. We report on the first optical polarimetric
observations of the Be/X-ray binary EXO\,2030+375. We find that the optical
($R$ band) light is strongly linearly polarized with a degee of
polarization of 19\%, the highest ever measured either in a classical or
Be/X-ray binary. We argue that the interstellar medium cannot account for
this high polarization degree and that a substantial amount must be
intrinsic to the source. We propose that it may result from  the
alignment of non-spherical ferromagnetic grains in the Be star disc
due to the strong neutron star magnetic field.

\end{abstract}


\begin{table*}
\begin{center}
\caption{Robopol observations of \exo. The polarization degree and angle 
 were calculated as $PD=\sqrt{q^2+u^2}$ and $PA=0.5\,{\tan}^{-1}(u/q)$.}
\label{pol}
\begin{tabular}{llccccccc}
\hline \hline \noalign{\smallskip}
Source	&Date		&JD		&Orbital	   &$q$  	    &$u$	       &$PD$		   &$PA$	      &$N_{\rm obs} \times t$\\
	&		&		&phase$^{\dag}$	   &		    &		       &(\%)		   &($^\circ$)        &(s)\\
\hline \noalign{\smallskip} 
\exo	&2013-10-06	&2456572.3914	&0.92		   &0.029$\pm$0.015 &0.193$\pm$0.015   &19.6$\pm$1.5	    &40.8$\pm$2.2   &$4\times150$\\
\exo	&2013-10-21	&2456587.3997	&0.25		   &0.026$\pm$0.009 &0.188$\pm$0.010   &19.0$\pm$1.0	    &41.0$\pm$1.4   &$6\times300$\\
\exo	&2013-10-29	&2456595.3656	&0.42		   &0.032$\pm$0.010 &0.182$\pm$0.010   &18.5$\pm$1.0	    &40.0$\pm$1.5   &$7\times200$\\
\exo	&2014-05-15	&2456793.5892	&0.73		   &0.046$\pm$0.071 &0.200$\pm$0.021   &21.3$\pm$2.0	    &35.2$\pm$2.6   &$6\times180$\\
\exo	&2014-06-22	&2456831.5212	&0.55		   &0.019$\pm$0.012 &0.184$\pm$0.011   &18.5$\pm$1.2	    &42.0$\pm$1.9   &$6\times125$\\
\exo	&2014-08-01	&2456871.5297	&0.42		   &0.022$\pm$0.021 &0.174$\pm$0.021   &17.6$\pm$2.1	    &41.4$\pm$3.3   &$5\times100$\\
\hline \hline \noalign{\smallskip}
\multicolumn{9}{l}{$\dag$: Orbital solution from \citet{wilson08}}\\
\end{tabular}
\end{center}
\end{table*}
\begin{table*}
\begin{center}
\caption{Robopol observations of four field stars in the vicinity of \exo.}
\label{polstar}
\begin{tabular}{llcccccc}
\hline \hline \noalign{\smallskip}
Source	&Date		&JD		&$q$		 &$u$		    &$PD$		&$PA$	    &Distance from\\
	&		&		&		 &		    &(\%)		&($^\circ$) &target ('')\\
\hline \noalign{\smallskip} 
Star 1	&2014-06-19	&2456828.5387	&0.014$\pm$0.009 &0.057$\pm$0.009   &5.9$\pm$0.9	&38$\pm$4   &57\\
Star 2	&2014-06-13	&2456822.4453	&0.008$\pm$0.005 &0.013$\pm$0.007   &1.5$\pm$0.7	&30$\pm$14  &90\\
Star 3	&2014-05-20	&2456798.5464	&0.009$\pm$0.009 &0.004$\pm$0.009   &1.0$\pm$0.9	&13$\pm$25  &24\\
Star 4	&2014-06-21	&2456830.4945	&0.058$\pm$0.018 &0.038$\pm$0.007   &7.0$\pm$1.5	&17$\pm$5   &40\\
\hline \hline \noalign{\smallskip}
\end{tabular}
\end{center}
\end{table*}

\section{Introduction}

Be/X-ray binaries (BeXB) are a subclass of high-mass X-ray binaries that
consist of a Be star and a neutron star \citep[see e.g.][]{reig11}. The
mass donor  is a relatively massive ($\simmore$ 10 $\msun$) and
fast-rotating ($\sim$80\% of break-up velocity) Be star, whose equator is
surrounded by a disc formed from photospheric plasma ejected by the star.
The equatorial disc is believed to be Keplerian and supported by viscosity
\citep{rivinius13}. The accreting component is a strongly magnetized
neutron star. Matter from the disc is transferred to the neutron star, 
latched onto the magnetic field lines and deposited onto a very small
fraction ($\sim$1\%) of the neutron star surface, the polar caps, where its
kinetic energy is converted into X-rays. 

Be stars are also observed as single objects, that is, not forming part of
a binary system \citep{porter03, rivinius13}. In principle, the underlying
Be star in a BeXB and in the isolated Be share the same physical properties
(mass, radius, luminosity, temperature) for the same spectral type.
However, there is strong evidence that the discs around Be stars in BeXB
are strongly affected by the neutron star, resulting in more compact and
denser discs \citep{okazaki02,reig11}.

The three observational properties that chartacterise Be stars are emission
lines (which give the name of the class, i.e. "e" stands for emission),
infrared excess and polarized light. Partially or totally filled-in
emission lines are typically the dominant feature in the optical and
infrared spectra of such stars. The shape and strength of the H$\alpha$
line are useful indicators of the state of the disc. Photometrically, Be
stars present redder colors than normal, i.e. non-emitting, B stars of the
same spectral type. Also, Be stars are brighter in the infrared than B
stars, an effect known as infrared excess. Optical polarized light at
1--3\% is a common feature among isolated Be stars \citep{yudin01}.

All these three phenomena are explained by radiation processes taking place
in the extended circumstellar disc. Emission lines result from continuous
hydrogen recombination of atoms excited by the radiation emitted from the
massive star. Similarly, the infrared excess is due to free-bound and
free-free radiation in the ionized circumstellar disc, with no evidence for
dust \citep{gehrz74,dachs88}. The continuum polarization is attributed to
Thomson scattering of unpolarized starlight in the disc. 
The maximum polarization level in an axisymmetric circumstellar
disc predicted by single-scattering plus attenuation models is about 2\%
\citep{waters92}. Although this value generally agrees well with the
observations \citep{yudin01}, the single-scattering calculation are too
simplistic as they assume optically thin discs. For gaseous discs with
extended optically thick regions, multiple scattering significantly
modifies the resultant continuous polarization. Monte Carlo simulations
allowing for multi-scattering show that the optical polarization level
increases as the optical depth increases, reaches a maximum around to
3--4\%, and then decreases for further increase of the optical depth
\citep{wood96,halonen13a}.

\exo\ is a BeXB that has now been observed for almost 30 years since its
discovery by {\it EXOSAT} in 1985 \citep{parmar89}. During this time, it
underwent two giant X-ray outbursts, in 1985 and 2006, and numerous 
orbitally-modulated ($P_{\rm orb}=46.0207\pm0.0004$ days) outbursts
\citep{wilson08}. The optical counterpart is a highly reddened V=19.6 B0Ve
star \citep{motch87,coe88}.

In this work, we present the results of the first polarimetric observations
of \exo\ in the optical band and report the discovery of an unexpected high
polarization degree. 

\begin{table}
\begin{center}
\caption{Photometric observations of \exo\ and nearby stars.}
\label{phot}
\begin{tabular}{@{~~}l@{~~}c@{~~}c@{~~}c@{~~}c}
\hline \hline \noalign{\smallskip}
Object	&$B$		&$V$		&$R$		&$I$		\\
\hline \hline \noalign{\smallskip}
\multicolumn{5}{c}{2011-08-27 ; JD 2,455,801.38} \\
\hline \noalign{\smallskip}
\exo	&$22.08\pm0.16$ &$19.31\pm0.03$	&$17.25\pm0.02$	&$15.16\pm0.02$ \\
Star 1	&$15.84\pm0.02$	&$14.97\pm0.02$	&$14.50\pm0.02$	&$14.04\pm0.02$	\\
Star 2	&$15.63\pm0.02$	&$14.76\pm0.02$	&$14.28\pm0.02$	&$13.81\pm0.02$	\\
Star 3	&$20.06\pm0.03$	&$18.27\pm0.02$	&$17.26\pm0.02$	&$16.34\pm0.02$	\\
Star 4	&$21.45\pm0.10$	&$18.62\pm0.02$	&$16.74\pm0.02$	&$15.01\pm0.02$	\\
 \hline \noalign{\smallskip}
\multicolumn{5}{c}{2014-07-17 ; JD 2,456,856.31} \\
 \hline \noalign{\smallskip}
\exo	&$21.86\pm0.10$ &$19.41\pm0.02$	&$17.31\pm0.01$	&$15.23\pm0.02$ \\
Star 1	&$15.85\pm0.01$	&$15.01\pm0.01$	&$14.53\pm0.02$	&$14.06\pm0.02$	\\
Star 2	&$15.64\pm0.01$	&$14.80\pm0.01$	&$14.31\pm0.02$	&$13.84\pm0.02$	\\
Star 3	&$20.07\pm0.03$	&$18.33\pm0.02$	&$17.28\pm0.02$	&$16.41\pm0.02$	\\
Star 4	&$21.30\pm0.08$	&$18.65\pm0.02$	&$16.76\pm0.02$	&$15.05\pm0.02$	\\
 \hline \noalign{\smallskip}
\multicolumn{5}{c}{2014-08-20 ; JD 2,456,890.52} \\
 \hline \noalign{\smallskip}
\exo	&$22.12\pm0.17$ &$19.43\pm0.03$	&$17.29\pm0.01$	&$15.29\pm0.03$ \\
Star 1	&$15.83\pm0.02$	&$15.02\pm0.02$	&$14.55\pm0.02$	&$14.09\pm0.03$	\\
Star 2	&$15.70\pm0.02$	&$14.81\pm0.02$	&$14.33\pm0.02$	&$13.86\pm0.03$	\\
Star 3	&$20.04\pm0.03$	&$18.32\pm0.02$	&$17.30\pm0.02$	&$16.37\pm0.03$	\\
Star 4	&$21.32\pm0.08$	&$18.65\pm0.02$	&$16.76\pm0.02$	&$15.09\pm0.03$	\\
\hline \hline \noalign{\smallskip}
\end{tabular}
\end{center}
\end{table}

\section{Observations}

Photometry and polarimetry were obtained with the 1.3 m telescope of the
Skinakas Observatory in Crete (Greece), which has a modified 
Ritchey-Chretien optical system with a 129 cm primary, 45 cm secondary, and
$f=7.54$.

\subsection{Polarimetric observations}

Polarimetric observations in the $R$ band were made with the Robopol
photopolarimeter (Ramaprakash et al. 2014, in preparation) attached to the
focus of the  telescope, on the nights indicated in Table~\ref{pol}.
Robopol is a 4-channel imaging photopolarimeter. The incident
light is split in two beams, each half incident on a half-wave retarder
followed by a Wollaston prism.   Every point in the sky is thereby
projected to four points on the CCD. The CCD chip is an ANDOR DW436
2048$\times$2048 with pixel size 13.5 $\mu$m, giving a field of view of $13'
\times 13'$. The photon counts, measured using aperture photometry, in each
spot are used to calculate the $U$ and $Q$ parameters of linear
polarization. The fast axis of the half-wave retarder in front of the first
prism is rotated by $67.5^{\circ}$ with respect to the other retarder. To
optimize the instrument sensitivity a mask is placed in the telescope focal
plane. The mask has a cross-shaped aperture in the center where the target
source is placed. The data were reduced following the pipeline procedure
described in \citet{king14}. In the instrument reference frame, the
horizontal channel measures the $u=U/I$ fractional Stokes parameter, while
the vertical channel measures the $q=Q/I$ parameter, simultaneously, with
a single exposure. 

In addition to the BeXB \exo, we also observed four nearby stars, which in
the target images are hidden behind the mask.  We chose these stars because
of their proximity to the target, their isolation from nearby stars that
could cause confusion, and because they cover all four directions around
the target. The results of the polarimetric observations are shown in
Table~\ref{polstar}.

\subsection{Photometric observations}

Johnson-Cousins-Bessel $BVRI$ photometric observations were made of the
field around \exo. We obtained the magnitudes of the target and the four
nearby stars for which polarimetry was obtained (Table~\ref{phot}). The
telescope was equipped with a  2048$\times$2048 ANDOR DZ436 CCD with a 13.5
$\mu$m pixel size (corresponding to 0.28 arcsec on the sky) providing a
field of view of 9.5 arcmin $\times$ 9.5 arcmin.    Reduction of the data
was carried out in the standard way using the IRAF tools for aperture
photometry. The photometry was accurately corrected for colour equations
and transformed to the standard system using nightly observations of
standard stars from Landolt's catalogue \citep{landolt92,landolt09}.
Usually, the error of the photometry is calculated as the root-mean-square
of the difference between the derived final calibrated magnitudes of the
standard stars and the magnitudes of the catalogue. However, because of the
faintness of the target and some of the nearby stars, we took the error to
be the larger of the root-mean-square and the uncertainty obtained from the
signal-to-noise ratio ($\sigma_B=1.086/SNR$).

\begin{figure}
\resizebox{\hsize}{!}{\includegraphics{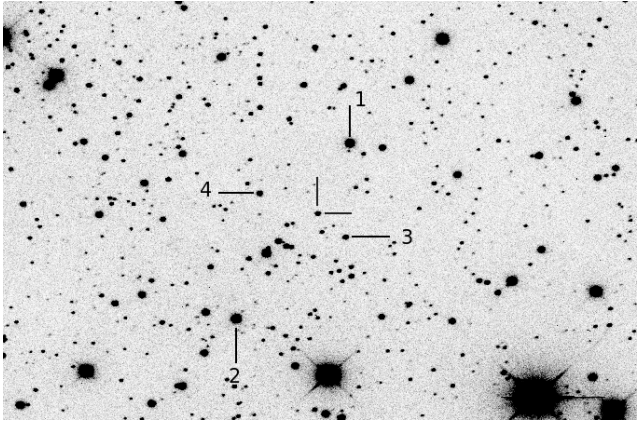} } 
\caption[]{Optical ($R$ band) field ($7.5\times4.2$) arcmin around the BeXB 
\exo. The four reference stars for which polarization was measured are 
indicated, together with the target.
 }
\label{field}
\end{figure}

\section{Results and discussion}

We performed polarimetric observations of the BeXB \exo\ at six different
epochs and found that the optical light in the $R$ band is highly linearly
polarized. No evidence for variability is observed neither in the degree of
polarization nor in the polarization angle, with weighted average values
of  $P(\%)=18.9\pm0.5$ and $\chi=40.4^{\circ}\pm0.8$, respectively . This
is the highest polarization level ever measured in a Be star. Before
attempting to find a mechanism to explain such a large polarization degree,
we must assess whether the interstellar medium (ISM) can account for the
measured polarization. Because the polarization of the interstellar medium
is closely related to the amount of extinction, we first estimate the color
excess of \exo\ from our photometry.

The observed colour of \exo\ is $(B-V)=2.6\pm0.1$
(Table~\ref{phot}), while the expected one for a B0V star is
$(B-V)_0=-0.29$ \citep{fitzgerald70,gutierrez-moreno79,wegner94}. Thus we
derive a colour excess of $E(B-V)=2.9\pm0.1$ or visual extinction $A_{\rm
V}=R \times E(B-V)= 9.0\pm0.3$, where the standard extinction law $R=3.1$
was assumed. In a Be star,  the total measured reddening is made up of two
components: one produced mainly by dust in the interstellar space along
the line of sight and another produced by the circumstellar gas around the
Be star. The  contribution of the circumstellar reddening to the color
excess $E(B-V)$ has been estimated in isolated Be stars to be $\simless
0.1$ mag \citep{dachs88}. To account for this extra reddening, we increased
the error in $E(B-V)$ to 0.2 mag. Thus we finally derive $E(B-V)=2.9\pm0.2$
mag. 

An important point to estimate the amount of polarization from the
ISM is the distance to the source. The distance to \exo\ is rather uncertain.
The first optical observations could only constrain the distance to be in
the range 2--7 kpc due to the unknown spectral type \citep{motch87}.  When
optical spectroscopy was performed and a confirmation that the optical
counterpart to \exo\ was indeed a Be star, then the maximum distance was
proposed to be less than 4 kpc, assuming an overestimated value of $M_V=-5$
\citep{janot88}. The distance derived from fits of various accretion torque
models to the X-ray luminosity versus spin rate relation agrees with the
range 4--5 kpc \citep{parmar89,reynolds96}. However, \citet{wilson02}
suggested a distance of $\sim$7 kpc using the relationship between
extinction and distance for objects in the Galactic plane.
Using our photometric data and taking the most recent
absolute magnitude calibration for a B0Ve star, $M_V=-3.9\pm0.5$
\citep{wegner06}, we find $d=7\pm3$ kpc. The error of 0.5 magnitudes in $M_V$
mainly accounts for the uncertainty in the spectral classification. The
error in the distance was obtained by propagating the errors in $E(B-V)$,
$V$, and $M_V$.

\subsection{Interstellar polarization}

Although we present the first polarimetric observations in the optical
band, it is not the first time that polarization has been measured in \exo.
\citet{norton94} performed a multi-wavelength analysis of \exo\ during a
normal outburst and reported on infrared polarimetric observations with the
IRCAM infrared camera in the $J$ and $K$ bands. They measured  average
polarization fraction of $10.0\pm0.6$\% in the J band and $4.2\pm0.2$\% in
the K band. \citet{norton94} suggested that the most likely origin of most of the
measured polarization is the interstellar medium, although an intrinsic
component of about 3\% in $J$ and 1\% in $K$ could not be ruled out.   They
based their conclusions on the average relationships between
interstellar polarization and extinction \citep{serkowski75,whittet78} and
the fact that interstellar polarization over 10\% in the J band has been
observed in OH/IR stars \citep{jones90}. 

In what follows, we present evidence that the ISM cannot
account for the entire observed optical polarizaton in \exo.

\begin{figure}
\resizebox{\hsize}{!}{\includegraphics{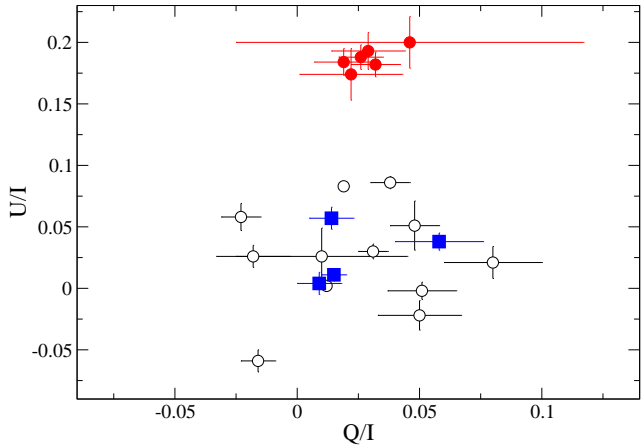} } 
\caption[]{$q-u$ diagram. Black open circles are field stars. Blue filled squares 
correspond to the four stars for which dedicated pointed observation were made. 
The \exo\ measurements are indicated by red filled circles. }
\label{qu}
\end{figure}

\begin{itemize}

\item {\it Polarization-extinction relationship}. The relationship between
polarization and extinction has been studied by a number of authors
\citep{hiltner56,serkowski75,jones89,fosalba02}. These studies show that
the polarization fraction increases as the extinction increases, albeit
with a large scatter. In fact, rather than a neat correlation, there is an
ensemble of points for the same or similar extinction. As the extinction
increases, the maximum also increases. 

\citet{hiltner56} found this limit in $P_{\rm max}/A_V=2.76$\%, similar to
that by \citet{serkowski75}, who found  $P_{\rm max}/E(B-V)=9$\%.  Note
that the data used to build these relations are based on stars with
$E(B-V)\simless 1$, hence relatively close to our Sun, and that the maximum
envelope represents optimum alignment efficiency, that is, it arises on
sight lines where the magnetic field is uniform, highly ordered, and where
the dust grains are completely alligned. The allignment and uniformity of
the magnetic field is expected to disappear as the line of sight crosses
more and more interstellar dust clouds. Thus the more distant objects are
expected to suffer from strong depolarization.  This effect is  apparent in
the $P_{\rm max}-E(B-V)$ plots \citep{serkowski75} or, equivalently the
$P_{\rm max}-A_V$ plots \citep{hiltner56}, where one can see that the
larger the extinction, the further away from the maximum line the star is.
It looks like there is a turn over at about 6--7\%. Using a much larger
sample of stars, \citet{fosalba02} found that the observed mean correlation
between extinction and ISM polarization is much smaller than what it is
expected from interstellar dust grains completely aligned under a purely
regular external magnetic field. Using their relation, $P_{\rm max,
ISM}(\%)=3.5\, E(B-V)^{0.8}$, the maximum contribution of the ISM to the
measured polarization towards \exo\ would be $P_{\rm ISM}=9$\%.
\citet{fosalba02} also found that the polarization degree as a function of
distance shows a maximum at $\sim2-3$\% for stars at 2--4 kpc and then
decreases slightly for more distant objects up to 6 kpc.

These are average relationships obtained by measuring the polarization
degree of a large number of stars in {\em all directions} of the Galaxy.
However, the polarization degree strongly depends on the galactic
longitude, with two maxima and two minima. The two minima are found at
50--70$^{\circ}$ and 230--250$^{\circ}$, while the two maxima at
130--140$^{\circ}$ and 320--330$^{\circ}$ \citep{krautter80,fosalba02}. The
galactic coordinates of \exo\ are $l=77.15^{\circ}$ and $b=-1.24^{\circ}$.
At that longitude, $P(\%)/A_v\sim1.0$ \citep{krautter80}. Taking
$E(B-V)=2.9$ mag and assuming $R_V=3.1$, the average polarization in the line
of sight to \exo\ would be $P_{\rm ISM}\sim9$\%.

\item {\it Planck's polarization maps}. The Planck survey has shown that
the polarization fraction in the galactic plane at 353 GHz is only a few
percent \citep[see Fig.4 in][]{planckXIX}. Moreover, the polarization
fraction decreases with increasing column density. There is a sharp decline
in polarization fraction starting at $N_H=2\times 10^{22}$ cm$^{-2}$, which
corresponds to $A_V=10$ mag, that is, approximately the extinction of \exo.
At this $A_V$, $p\sim5$\%. The decrease of polarization fraction towards
large column densities is interpreted as being due to depolarization, that
is, a gradual loss of alignment of dust grains or to fluctuations in the
orientation of the magnetic field along the line of sight. Note that this
$p$ refers to the polarization fraction at 353 GHz. However, the
submillimitre polarization intensity, $P_S$, at 353 GHz is related to the
optical polarization fraction in $V$, $p_V$, through
$R_{P/p}=P_S/p_V=5.6\pm0.4$ MJy sr$^{-1}$ \citep{planckXXI}. In the
direction of \exo, $\log P_S$ is estimated to be $\log P_S \sim -(1-0.3)$
MJy sr$^{-1}$ \citep[see Fig. 2 in][]{planckXIX}. Thus, the polarization
fraction due to the ISM, according to {\it Planck} results, is again
$p_V\simless 9$\%.

\item {\it Field stars}.  We measured the $q$ and $u$ Stokes parameters of
12  stars in the Robopol field of view not blocked by the mask in the
target images. To reduce the uncertainty in the Stokes  parameters of these
stars, we combined several images when \exo\ was at the center of the
instrument mask. The values were median filtered (averaged in case of 2)
measurements of the same star. The main criterion for the selection of
these stars was to have the four light spots free of nearby sources to
avoid overlapping, signal-to-noise ratio in flux larger than 10, and the
change in the polarization angle is lower than 30$^{\circ}$. 
Figure~\ref{qu} shows the $q-u$ plane of the field stars and the target.
The four field stars for which dedicated observations were obtained (see
Table~\ref{pol}) are also displayed with blue filled squares. Of all the
sources, \exo\ (filled red circles) occupies  the most distant position
from the origin, indicating that it is the star with the highest
polarization degree. Some field stars also display non-zero values of the
Stokes parameters, but all the stars have $P_R\simless9$\%, that is, the
measured polarization degree of the stars in the vicinity of \exo\ agrees
very well with what it is expected from the polarization-extinction
relationship. 

\item {\it Wavelength dependence of the ISM polarization}. The spectrum of
the ISM polarization  obeys a well known empirical law given by
$P(\lambda)/P_{\rm max}=\exp[-k \ln^2(\lambda_{\rm max}/\lambda)]$
\citep{serkowski75}. It peaks in the optical band and falls both in the UV
and near IR regions. $\lambda_{\rm max}$ is the wavelength at which $P_{\rm
max}$ is observed and $k$ is a measure of the sharpness of the spectrum.
$\lambda_{\rm max}$ has been seen to correlate with the total to selective
galactic extinction $R_V$ as \citep[see e.g.][]{whittet78}

\begin{equation}
\label{ser}
R_V=A_V/E(B-V)=(5.6\pm0.3)\,\lambda_{\rm max}
\end{equation}

\noindent where $\lambda_{\rm max}$ is given in $\mu$m . The presence of an
intrinsic polarization component can be inferred if the wavelength
dependence of the ISM polarization is different from that of the
established curve related to interstellar effects. Because the polarization
from the interstellar medium is supposed to be constant with time, we can 
use the results in the $J$ and $K$ bands from \citet{norton94} together
with our observation in the $R$ band to estimate the total to selective
extinction from eq.~(\ref{ser}).  The aim is to check whether the
polarimetric data gives a value of $R_V$ consistent with the average
galactic value of $R_V=3.1$ \citep{cardelli89}. Assuming that 
$P(R)=19.0\pm0.5$\%, $P(J)=10.0\pm0.6$\%, and $P(K)=4.2\pm0.2$\% are
entirely caused by the ISM, then 

\begin{equation}
\label{rj}
\frac{p_R}{p_J}=\frac{exp[-k\, \ln^2(\lambda_{\rm
max}/\lambda_R)]}{exp[-k \, \ln^2(\lambda_{\rm max}/\lambda_J)]}
\end{equation}

\noindent and equivalently for the K band

\begin{equation}
\label{rk}
\frac{p_R}{p_K}=\frac{exp[-k\, \ln^2(\lambda_{\rm
max}/\lambda_R)]}{exp[-k\, \ln^2(\lambda_{\rm max}/\lambda_K)]}
\end{equation}

\noindent Dividing equation \ref{rj} by \ref{rk} and solving for
$\lambda_{\rm max}$ we obtain $R_V=0.77^{+0.75}_{-0.41}$, with a  $3\sigma$
upper limit of $R_V=2.62$, which is not consistent with the standard
ratio of total to selective extinction of $R_V=3.1$ \citep{cardelli89}. Thus,
either the extinction toward \exo\ is peculiar and does not obey the
average galactic extinction law or the observed polarization cannot be
explained by the ISM.

\end{itemize}

Taking all the above considerations into account, we conclude that the
ISM in the direction of \exo\ cannot explain the observed polarization degree.

\subsection{Polarization in \exo}

Models that attempt to explain the polarization observed in Be stars as
Thomson scattering predict maximum polarization fraction of 3--4\%. 
Observations of classical Be stars agree with this value \citep[see
e.g.][]{yudin01}.  From the previous discussion we conclude that it is
extremely unlikely that the interstellar medium can account for the
measured polarization in \exo. With the available data, it is difficult to 
separate the interstellar contribution from the observed polarization. As a
rough approximation, we can assume that the measured polarization of the
field stars is entirely of  interstellar origin. We then calculate the
average polarization degree of the nearby stars and subtract it vectorially
from the total observed polarization \citep[see e.g.][]{clarke10}. The
result is that $\sim$16\% must be intrinsic. Such a high degree of
polarization implies that a mechanism other than Thomson scattering might
be at work in \exo. 

The main difference between an isolated Be star and a BeXB is the presence
of a neutron star with a strong magnetic field. One is tempted to attribute
the high polarization degree to this magnetic field. The alingment of
non-spherical iron grains by the neutron star magnetic field
might be an alternative mechanism to explain the high polarization in \exo.
The presence of iron in the Be discs is well documented. High resolution
optical spectra of Be stars reveal the presence of Fe II emission lines
with the strongest transitions in the optical band at $\lambda$5169 \AA,
$\lambda$4584 \AA, and $\lambda$5317 \AA\ \citep{hanuschik96}. In addition,
BeXB in general and \exo\ in particular, show strong ($EW \sim 0.18$ keV)
emission line in the X-ray spectra at 6.4-6.6 keV
\citep{reynolds93,reig99}. This line is interpreted as reprocessing of the
hard X-ray continuum in relatively cool matter. Near-neutral iron generates
a line centered at 6.4 keV, while this energy increases as the ionization
stage increases. Therefore the detection of an iron line implies the
presence of relatively cold matter in the vicinity of the X-ray source,
which in the case of BeXB is the circumstellar disc.

 For the proposed mechanism to work, two conditions have to be met: {\em
i)} temperatures below the iron condensation\footnote{The condensation
temperature of an element is defined as the temperature at which half of
that element is removed from the gas phase due to the formation of solid
condensates \citep{savage92}.} critical value are needed and {\em ii)} the
neutron star has to physically interact with the disc. 

The condensation temperature of iron is $\sim 1300$ K \citep{savage92}.
Above this temperature no grains can be formed  because refractory elements
like Fe cannot condense out of the gas phase. Models of the temperature
structure of viscous discs in Be stars show that those discs are highly
nonisothermal, mainly in the inner denser parts \citep{carciofi06}. When
the density is large enough ($\rho_0\simmore 5 \times 10^{-12}$ g
cm$^{-3}$) the disc develops a cool equatorial region close to the star. As
the density increases, the minimum temperature decreases and moves further
away from the central star \citep{carciofi08,sigut09}. Still the lowest
temperature predicted by the model for a B0 star is 6000--8000 K, well
above the temperature for grain condensation of iron. However, these
simulations predict the temperature profile relatively close to the star,
whereas the temperature in the outermost regions, where the disc meets the
interstellar space, is expected to be significantly lower and so is the
particle density content.

A dipolar magnetic field decreases with distance as $B=B_0(R_{\rm
NS}/r)^3$, where $B_0$ is the neutron star polar magnetic field. This
pronounced decrease of the strength of the magnetic field with distance
implies that the disc feels a substantial magnetic field only when the
radius of the disc is comparable to the periastron distance.  The H$\alpha$
equivalent width of \exo\ varies in the range 8-20 \AA\
\citep{norton94,reig98,baykal08}, which implies a disc radius of the order
of $R_{\rm disc}\simmore 10 R_*$ \citep{grundstrom06,carciofi11}. Also,
Monte Carlo simulations that attempt to reproduce the continuum
polarization in equatorial discs show that polarization is formed far away
from the star. Approximately 95\% of the maximum polarization is reached
only when the disc size is about ten stellar radii \citep{carciofi11}.
Assuming a stellar radius of  $R_*\approx 8 \rsun$ for a B0V star, the disc
radius would be $R_{\rm disc}\sim 5.5\times 10^{12}$ cm, which roughly
coincides with the periastron distance using the orbital solution of
\citet{wilson08} and assuming a stellar mass $M_*=16 \msun$. 

Further evidence for close encounters between the neutron star and the disc
during periastron comes from the X-rays.  \exo\ is unique among other BeXBs
in that it shows stable and long-lasting periodic X-ray type I outbursts
\citep{wilson08}. This suggests that  the neutron star physically interacts
with the disc at every periastron passage. With a typical magnetic field of
$\sim 10^{12}$ G \citep{reig99,wilson08}, the alignment efficiency is very
high and the iron grains get magnetized very fast.  Because the iron grains
are inmersed in a gas, gas-grain collisions tend to restore random
orientation. Thus, whether the alignment survives throughout the orbital
period depends on the relaxation time, which in turn depends on the size of
the grains and the temperature and particle density of the gas \citep[see
eq.~(4.26) in][]{whittet92}.  For the typical parameters expected in a Be
star disc, the relaxation time-scales are of the order of tens of days to
years. In this scenario, we would expect to observe orbital polarimetric
variability when the relaxation time is smaller than the orbital period.
The lack of orbital variability in \exo\ indicates that the continuous
passages of the neutron star through periastron ensures that the grain
alignment in the outer parts of the disc is not lost.   

A prediction of the model is that polarization and X-ray activity must
correlate in BeXBs, in the sense that a high degree of polarization should be
observed during type I outbursts. 

\section{Conclusion}

We have performed optical ($R$ band) polarimetric and photometric
observations of the Be/X-ray binary \exo\ and found the highest optical
polarization degree ever reported in a Be star. Over a year the source has
shown a fairly constant level of polarization at $\sim19$\% and
polarization angle of 40$^\circ$.  We have shown evidence indicating that
the contribution of the interstellar medium cannot account for the total
measured polarization, leaving room for a large intrinsic contribution. 
This evidence is based on the inconsistent value of the total to selective
extinction $R_V$ obtained assuming that {\it all} the measured polarization
is due to the ISM, the relationship between polarization and extinction in
the ISM, and the measured polarization of stars in the vicinity of the
target. The origin of the polarized light in \exo\ is not clear, but a
different mechanism from that proposed in classical Be stars must be
invoked.

\section*{Acknowledgments}

RoboPol is a collaboration involving the University of Crete, the
Foundation for Research and Technology - Hellas, the California Institute
of Technology, the Max-Planck Institute for Radioastronomy, the Nicolaus
Copernicus University, and the Inter-University Centre for Astronomy and
Astrophysics. This work was partially supported by the ``RoboPol'' project,
which is implemented under the ``Aristeia'' Action of the  ``Operational
Programme Education and Lifelong Learning'' and is co-funded by the
European Social Fund (ESF). We also acknowledge support from the COST
Action MP1104 ``Polarisation as a tool to study the Solar System and
beyond". K.T. acknowledges support by FP7  Marie Curie Career Integration
Grant PCIG-GA-2011-293531 ``SFOnset" and the EU FP7 Grant
PIRSES-GA-2012-31578 ``EuroCal".

\bibliographystyle{./mn2e}
\bibliography{../../../pol}

\end{document}